\documentclass{article}
\usepackage{spconf,amsmath,graphicx}
\usepackage{multirow}
\usepackage{multicol}
\usepackage{tipa}
\PassOptionsToPackage{hyphens}{url}\usepackage{hyperref}
\usepackage{todonotes}

\title{INDEPENDENT AND AUTOMATIC EVALUATION OF ACOUSTIC-TO-ARTICULATORY INVERSION MODELS}
\name{Maud Parrot$^{1}$, Juliette Millet$^{1,2,3}$, Ewan Dunbar$^{1,2}$}
\address{$^1$ CoML, ENS/CNRS/EHESS/INRIA/PSL Research University, Paris, France\\$^2$ Universit{\'{e}} de Paris, LLF, CNRS, Paris, France\\$^3$CRI, Département Frontières du Vivant et de l'Apprendre, IIFR, Universit{\'{e}} de Paris}
\begin{document}
%\ninept
%
\maketitle
\begin{abstract}

Reconstruction of articulatory trajectories from the acoustic speech signal has been proposed for improving speech recognition and text-to-speech synthesis. However, to be useful in these settings, articulatory reconstruction must be speaker independent. Furthermore, as most research focuses on single, small datasets with few speakers, robust articulatory reconstrucion could profit from combining datasets. Standard evaluation measures such as root mean square error and Pearson correlation are inappropriate for evaluating the speaker-independence of models or the usefulness of combining datasets. We present a new evaluation for articulatory reconstruction which is independent of the articulatory data set used for training: the \textit{phone discrimination ABX task}. We use the ABX measure to evaluate a Bi-LSTM based model trained on 3 datasets (14 speakers), and show that it gives information complementary to the standard measures, and enables us to evaluate the effects of dataset merging, as well as the speaker independence of the model.

\end{abstract}
\section{Introduction}
\label{sec:intro}

Acoustic-to-articulatory inversion is the problem of finding a mapping from  acoustic features to a set of articulatory measures (see  \cite{Deep_rnn_arti_inversion, 30_yu2018synthesizing,28_tobing2017deep,26_mitra2017joint,13_seneviratne2019multi,9_porras2019dnn} for recent models: see \cite{6_richmond2015use} for a review). Reconstructed articulatory trajectories have been shown to improve text-to-speech synthesis \cite{24_cao2017integrating,6_richmond2015use}, speech accent conversion \cite{accent_conversion}, and automatic speech recognition \cite{speech_reco_using_arti,26_mitra2017joint}, in particular for dysarthric speech \cite{5_yilmaz2019articulatory}; they can also be used in the automatic detection of clinical conditions which have an impact on speech production, such as Parkinson's  \cite{25_hahm2015parkinson}. \emph{Speaker independent} reconstruction is essential for most of these applications, and, as such, some method for evaluating speaker-independent articulatory reconstruction is needed.

The two principal metrics for evaluating articulatory reconstruction are the root mean square error (RMSE) and the Pearson correlation coefficient (PCC)  between the measured and the predicted articulatory trajectories. However, the goal of a speaker-independent model is not precise prediction of reference measures. First, the shapes of speakers' vocal tracts vary in ways that cannot be captured by simple normalization. Second, recording conditions (coil placement in electromagnetic articulography: EMA), vary across and within recordings. Moreover, acoustic-to-articulatory inversion is a one-to-many problem: two different articulatory trajectories can produce the same sound. We do not want to penalize models that choose to reconstruct another trajectory which is correct but different from the reference. Finally, articulatory data is needed to calculate these measures, which is costly to obtain.

We propose a standardized evaluation based on an \textit{ABX phone discrimination test} \cite{schatz2013evaluating} of trajectories reconstructed for an acoustic-only corpus. The evaluation has the advantage of being \emph{independent} of the training data set and of the true reference articulatory trajectories, much like the independent evaluation of \cite{7_richmond2013evaluation}, which uses human listeners to evaluate speech re-synthesis on the basis of the reconstructed trajectories. Our evaluation, however, is completely \emph{automatic}.\footnote{All code for pre-processing the datasets and for training and testing the model is fully available at \url{https://github.com/bootphon/articulatory_inversion.git}}

We train a bi-LSTM model closely resembling that of \cite{Deep_rnn_arti_inversion} on three data sets (MOCHA--TIMIT, EMA--IEEE, USC--TIMIT), and apply the ABX phone discrimination evaluation on different training set, validation set, and test set. We compare ABX phone discrimination scores to standard metrics.

\section{Method}
\label{sec:method}
\subsection{Model architecture}
\label{sec:model}
\begin{figure}[htb]
  \centering
  \centerline{\includegraphics[width=5cm]{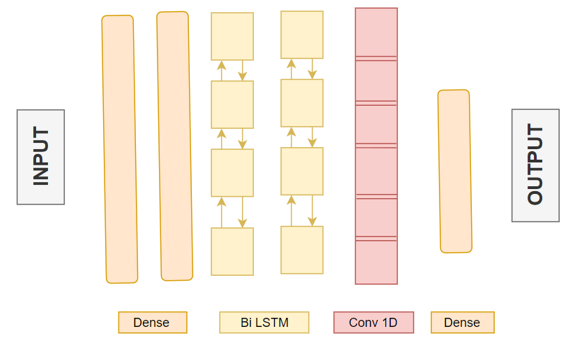}}
\hfill
\label{fig:archi_schema}
\caption{Neural network architecture used in this paper.}
\end{figure}

To demonstrate our evaluation, we use a bidirectional recurrent neural network architecture similar to that of \cite{Deep_rnn_arti_inversion}, but with the addition of a convolutional layer that acts as a low pass filter after the readout layer. The network has two feed-forward layers of 300 units each that act as feature extractors, two bidirectional layers of 300 units each, a convolutional layer as a low pass filter, and, finally, a feed-forward layer with as many units as the number of trajectories predicted (see Figure \ref{fig:archi_schema}).

\subsection{Loss function}
The usual loss function for articulatory inversion is the root mean square error (RMSE), which minimizes the L2 distance between the real and predicted trajectory.
The Pearson correlation coefficient (PCC) is also typically examined, in addition to the RMSE values, as it ignores any systematic differences between speakers that can be described linearly. The PCC  measures the degree of linear relationship between two variables $X$ and $Y$.
\begin{align}
    PCC(X,Y) = \frac{Cov(X, Y)}{\sigma_{X}\sigma_{Y}}
\end{align}

We experiment with including the PCC in the loss function, $y$ the reconstructed trajectory $y^*$ the reference:
\begin{equation}\label{eq:loss_comb}
    \mathcal{L}(y,y^*) = RMSE(y,y^*) - \beta\cdot PCC(y,y^*) 
\end{equation}
We use $\beta=1000$ (in order to account for the differences in scale between the RMSE and PCC values).\footnote{We ran experiments varying the weights of the RMSE and the PCC, but found it had no systematic effect on the results.}

\subsection{Convolutional layer as low pass filter}
    
 Articulatory gestures are not only continuous but also smooth. We impose smoothing on the predicted output trajectories by integrating a convolutional layer that acts as a low pass filter at the output of our neural network. This avoids unnecessary backpropagation of error due to non-smooth predicted trajectories. We used the following filter of order 5, with a Hanning window to restrict the support, where $N$ is the size of the Hanning window and $f_t$,  $\forall n \in [0,N-1]$:
 
 \small
 \begin{align}\label{eq:weights_filter}
w(n) & \propto & \Big(1- cos(2 \pi \frac{n}{N-1}) \Big)sinc\Big(2 \pi f_t (n-\frac{N-1}{2})\Big)  
 \end{align}
 \normalsize

 \subsection{ABX phone discrimination evaluation}
An \textit{ABX phone discrimination test} of a representation of speech consists of extracting the representations of triplets of stimuli (A, B, and X), and computing the distance $d(\mbox{A},\mbox{X})$, between A and X, and $d(\mbox{B},\mbox{X})$, between B and X. X is of the same phonetic category as either A or B. Taking A to be the correct answer, we compute $\delta = d(\mbox{B},\mbox{X}) - d(\mbox{A},\mbox{X})$.  If $\delta>0$, the model has chosen A; if $\delta<0$, it has chosen B. As in previous work evaluating acoustic models with this method \cite{versteegh2015zero,dunbar2017zero}, the percent correct for all pairs of categories are combined into a global ABX discriminability score. We follow previous literature and use dynamic time warping (DTW) based on cosine distance to align sequences of differing length \cite{senin2008dynamic}. The final distance between the two sequences is the mean of the cosine distance between the matched frames along the alignment path.

\section{ARTICULATORY DATASETS}
\label{sec:datasets}
    \subsection{Description}
 
All the data used are freely available. MOCHA--TIMIT\footnote{\url{http://data.cstr.ed.ac.uk/mocha}} is a database that contains EMA and acoustic data for 460 utterances (20 min) read by two English speakers \cite{description_mocha};  
USC--TIMIT\footnote{\url{https://sail.usc.edu/span/usc-timit/}} \cite{usc_timit_description} provides EMA data for four speakers on the MOCHA--TIMIT sentences (15 min); and EMA--IEEE\footnote{\url{https://yale.app.box.com/s/cfn8hj2puveo65fq54rp1ml2mk7moj3h/folder/30415804819}} \cite{Haskins} contains eight speakers reading 720 sentences, once each at a normal rate, and once at a fast speech rate (47 min per speaker).\footnote{We also conducted held-out tests on the single-speaker MNGU0 database, which we do not report due to preprocessing issues.} The combined duration is 461 min.

\subsection{Articulatory trajectories}
    
We use measures in the sagittal plane ($x$: back to front of head; $y$:  chin to  forehead). The two-dimensional articulatory points available in our datasets are: tongue body (TB), tongue tip (TT), tongue dorsum (TD), upper lip (UL), lower lip (LL), lower incisor (LI), and velum (V).
Not all measures are available for all speakers. The velum trajectory is only available in MOCHA--TIMIT, and we exclude specific articulators for certain speakers where the standard deviation was less than 0.5~mm and visual verification suggested strongly that the measure was wrong.\footnote{The list of articulators used by speaker is available at \url{https://github.com/bootphon/articulatory_inversion.git}} In training conditions combining speakers within corpora, we use the common articulators.

\subsection{Vocal tract parameters}
\label{sec:vocal_tract}
As in previous works \cite{1_speaker_adaption_vtln,29_sivaraman2016vocal}, we add vocal tract variables, using slightly different formulas from \cite{1_speaker_adaption_vtln}. We calculate two tract variables from the position of the lips, the vertical lip aperture (VLA) and the horizontal lip protrusion (HPRO):

{\centering
\begin{minipage}{.48\linewidth}
\begin{equation}
\mbox{VLA} = \mbox{UL}_y - \mbox{LL}_y
\end{equation}
\end{minipage}%
\begin{minipage}{.5\linewidth}
\begin{equation}
\mbox{HPRO} = \frac{\mbox{UL}_x + \mbox{LL}_x}{2}
\end{equation}
\end{minipage}\vspace{\baselineskip}\par
}

We also add the tongue tip constriction (TTC: the cosine of the angle of the tongue tip off the horizontal axis) and the tongue body constriction (TBC: the cosine of the angle of the tongue body off the horizontal axis).

{\centering\noindent
\begin{minipage}{.5\linewidth}
\begin{equation}
  \mbox{TTC} = \frac{\mbox{TT}_x}{\sqrt{\mbox{TT}_x^2 + \mbox{TT}_y^2}}
\end{equation}
\end{minipage}%
\begin{minipage}{.5\linewidth}
\begin{equation}
  \mbox{TBC} = \frac{\mbox{TB}_x}{\sqrt{\mbox{TB}_x^2 + \mbox{TB}_y^2}}
\end{equation}
\end{minipage}\par
}

\section{Experiments}
\label{sec:experiments}
% First general explanation

We compare standard reconstruction scores against ABX scores as evaluations of speaker-independence, both within and across corpora. 
Within corpus, we compare models trained in a  \textbf{speaker-specific} setting, on a single speaker, with models trained in a \textbf{speaker-independent} setting, training on multiple speakers. We randomly hold out data in the speaker-specific setting for validation and test (for calculating the reconstruction scores: 70\% train, 10\% validation, 20\% test). In the speaker-independent setting, we validate on a subset of the speakers, test on one speaker, and train on the rest. The speaker-specific model gives an expected upper bound on reconstruction. We expect that speaker-independent models will give poorer reconstruction, but we seek to use the ABX score to assess whether this degradation is due to failure to reconstruct linguistically relevant articulatory information, or failure to reconstruct speaker-specific detail.  

To test the effect of merging corpora, we compare a model trained in a \textbf{multi-corpus} setting (with speakers EMA--IEEE:~M01, MOCHA--TIMIT:~FSEW0, and USC--TIMIT:~M1 held out for validation and test) 
against a \textbf{single-corpus} setting, training on EMA--IEEE, which contains the most complete set of articulators (speaker M01 still hold out for test). Here, rather than training only on articulators common to all speakers, we learn to reconstruct all trajectories by ignoring error on missing articulators for backpropagation.

\subsection{Model parameters}

We use the Adam optimizer with early stopping on the validation set (learning rate 0.001, batch size 10,  patience  5). The weights of the low pass filter are fixed according to \eqref{eq:weights_filter} with $N=50$ to give a transition band of 0.08. The convolution has one channel, stride of 1, and padding such that the output has the same size as the input. The cutoff frequency $f_c$ is 10Hz.

\subsection{Data preprocessing}

We use as input the thirteen first MFCCs + $\Delta$ + $\Delta \Delta$ with window size of 25ms and stride of 10ms. We  add 10 context windows: the 5 previous and 5 following frames, as in \cite{deep_archi_arti_inv}. We remove silences based on the transcription file (when available). We normalize the MFCCs per speaker, removing the mean and dividing by the standard deviation.

We pre-smooth the articulatory trajectories, applying a low pass filter with a cutoff frequency of 10Hz for all the data sets except for EMA--IEEE, for which we use 20Hz. We remove leading and trailing silences ends using the transcription when available. We reduce EMA sampling rates to 100Hz to have a single articulatory frame per MFCC frame. Since  EMA coils move gradually during recording \cite{These_richmond}, we normalize each articulatory measure by subtracting the mean over the 60 previous and following recordings of the same speaker and divide by the speaker-specific standard deviation.

\subsection{Reconstruction scores}
\label{sec:reconstruction-evaluation}

RMSE (mm) and RMSE computed on normalized trajectories are computed on every feature except TTC and TBC, and PCC is computed on every feature available.

\subsection{ABX scores}
\label{sec:abx-evaluation}

The ABX test is performed on the 1-second English (speech-only) test data set from the Zero Resource Speech Challenge 2017 \cite{dunbar2017zero}, consisting of data from 24 speakers taken from the LibriVox audio book collection, labelled using the 39 CMUDICT phonemes plus \texttt{ax} for [\textipa{@}]. Stimuli are triphones differing in the central phone (\emph{beg}--\emph{bag}, \emph{api}--\emph{ati}, etc). \textit{Within-speaker} triplets contain three triphones from a single speaker (e.g., $A=\textrm{beg}_{T1}$, $B=\textrm{bag}_{T1}$, $X=\textrm{bag}'_{T1}$). In \textit{across-speaker} triplets, $A$ and $B$ come from the same speaker, and $X$ to another. $A=\textrm{beg}_{T1}$, $B=\textrm{bag}_{T1}$, $X=\textrm{bag}_{T2}$. The scores for a given contrast are first averaged across all (pairs of) speakers for which triplets can be constructed, before averaging over all contexts, and over all pairs of central phones and being converted to an error rate by subtracting from $1$.  We exclude contrasts with less than three contexts and for which critical articulators were missing from the data.\footnote{For example, oral--nasal contrasts such as [ana]--[ada]--[ana], which depend necessarily on the position of the velum: a complete list is provided at \url{https://github.com/bootphon/articulatory_inversion.git}}

\section{Results}
\label{sec:results}

The model attains  reconstruction scores on speaker-specific training  for FSEW0 which are comparable to existing results (RMSE-mm: 1.43, RMSE-norm: 0.55, PCC: 0.77).

\begin{table*}[h!]
\centering

\begin{tabular}{l|p{7mm}|p{7mm}|p{7mm}|p{7mm}|p{7mm}|p{7mm}|p{7mm}|p{7mm}|p{7mm}|p{7mm}|p{7mm}|p{7mm}|p{7mm}|p{7mm}|p{7mm}|}
\cline{2-16}
  & \multicolumn{5}{|c|}{\textbf{MOCHA--TIMIT}} & \multicolumn{5}{|c|}{\textbf{USC--TIMIT}} & \multicolumn{5}{|c|}{\textbf{EMA--IEEE}} \\
\cline{2-16}
 & \multicolumn{1}{|c|}{R} & \multicolumn{1}{|c|}{Rn} & \multicolumn{1}{|c|}{PCC} & \multicolumn{1}{|c|}{A-w} & \multicolumn{1}{|c|}{A-a} & \multicolumn{1}{|c|}{R} & \multicolumn{1}{|c|}{Rn} & \multicolumn{1}{|c|}{PCC} & \multicolumn{1}{|c|}{A-w} & \multicolumn{1}{|c|}{A-a} & \multicolumn{1}{|c|}{R} & \multicolumn{1}{|c|}{Rn} & \multicolumn{1}{|c|}{PCC} & \multicolumn{1}{|c|}{A-w} & \multicolumn{1}{|c|}{A-a} \\\hline
\multicolumn{1}{|l|}{\textbf{Sp}} & 1.380 & 0.557 & 0.759 & 23.9 & 32.2 & 1.478 & 0.608 & 0.747 & 24.5 & 33.8 & 1.557 & 0.501 & 0.840 & 22.1 & 30.5 \\\hline
\multicolumn{1}{|l|}{\textbf{Ind}} & 2.184 & 0.851 & 0.417 & 24.6 & 32.0 & 2.310 & 0.917 & 0.199 & 24.3 & 33.9 & 2.198 & 0.688 & 0.672 & 18.4 & 24.8\\\hline
\end{tabular}
\caption{ {Comparison between speaker-specific (\textbf{Sp}) and speaker-independent (\textbf{Ind}) settings. R: RMSE, Rn: normalized RMSE, A-w: within-speaker ABX, A-a: across-speaker ABX. Smaller is better for all measures except PCC.  Scores are averages across training subsets (see section \ref{sec:experiments}).}}
\label{tab: results_global}
\end{table*}

\begin{table*}[h!]
\centering
\begin{tabular}{c|c|c|c|c|c|c|c|c|c|c|c|}
\cline{2-12}
& \multicolumn{3}{c|}{\textbf{M01 (EMA--IEEE)}} & \multicolumn{3}{c|}{\textbf{M1 (USC--TIMIT)}} & \multicolumn{3}{c|}{\textbf{FSEW0 (MOCHA)}} &
\multicolumn{2}{c|}{\textbf{ABX}} \\ \cline{2-12} 
 & R          & Rn         & PCC       & R          & Rn        & PCC       & R           & Rn         & PCC         
 & w& a\\ \hline

\multicolumn{1}{|c|}{\textbf{Single-corpus}}     & 1.79       & 0.66       & 0.72      & 2.05       & 1.13      & 0.02      & 2.72        & 1.11       & 0.08         
&      18.9       &    25.0        \\ \hline
\multicolumn{1}{|c|}{\textbf{Multi-corpus}}         & 1.80       & 0.66       & 0.71      & 1.89       & 1.04      & 0.14      & 2.61        & 0.98       & 0.22       
&   19.7          &       26.7     \\ \hline
\end{tabular}
\caption{ {Effects of training one or multiple corpora.}}
\label{tab:merge}
\end{table*}

\subsection{Speaker-independence within corpus}

Table \ref{tab: results_global} compare the average across speakers  in the speaker-specific setting against the average over all one-speaker-held-out training configurations in the speaker-independent setting.

As expected, the speaker-independent condition shows degradation in the three reconstruction scores, compared to the
speaker-specific condition.

The ABX
scores, calculated on an external speech corpus, provide a different 
picture. The differences between the two training conditions are 
small for both MOCHA--TIMIT and USC--TIMIT. 
Although we lose information about detailed articulatory tracks 
in the speaker-independent condition, this information is not relevant
to coding phonemic contrasts. 
In the EMA--IEEE corpus, the loss in reconstruction  between speaker-specific and speaker-independent conditions is smaller. This may be due to having more data per speaker, or it may be that the measures in this corpus are more similar across speakers. The ABX scores, however, show an improvement in the speaker-independent condition, meaning that, with this corpus, the speaker-independent model not only does not lose linguistically relevant information, but in fact even better reconstructs linguistically relevant articulatory information. This fact is lost by looking only at the reconstruction scores.

 \subsection{Merging corpora}
  
Results comparing multi-corpus with single-corpus training are shown in Table \ref{tab:merge}.
ABX scores in the \textbf{multi-corpus} condition are similar to those in the \textbf{single-corpus} condition, and, in fact, slightly worse, suggesting that adding speakers from additional corpora to the training is not beneficial to the speaker-independent model. Again, the reconstruction measures do not go in the same direction, improving slightly when corpora are added. It is possible that the addition of more data is helping to better reconstructing (informative) speaker-specific subphonemic information; but we observe that the appreciable improvement is exclusively in the novel corpora, suggesting that the improvements in reconstruction may be due to improved modelling of acoustic channel or coil placement properties specific to these corpora. This points to the differences among corpora, and the incompatibility that can exist between them, leading to potentially worse articulatory reconstructions when using them together, information that would be lost without the ABX score.

\section{Conclusion}
\label{sec:conclusion}

We have proposed an ABX phone discrimination measure for the evaluation of speaker-independent acoustic-to-articulatory models. The measure is independent of the articulatory trajectories, and thus does not penalize models for failing to capture speaker-specific articulatory details, it comes from a single external corpus, avoiding the inherent instability of held-out measures, and it is automatic, unlike speech-synthesis based evaluations.  
 Our ABX score only assesses the presence of information needed to contrast the phones labelled in the corpus used (40 English phoneme labels), but they can be replaced by finer-grained allophonic labels, if desired. One caveat of phone discriminability is that it will vary as a function of the set of articulatory dimensions reconstructed, not only of how well they are reconstructed. So this measure can only be used to compare models that reconstruct the same articulators. Nevertheless, we have shown that it can give important information complementary to traditional reconstruction scores, indicative of the degree to which improvements or declines in reconstruction are due to failure to reconstruct speaker-specific properties.
 
 \section{Acknowledgements}
This research was supported by the \'Ecole Doctorale Frontières du Vivant (FdV) -- Programme Bettencourt, by Facebook AI Research, and by grants ANR-17-CE28-0009 (GEOMPHON), ANR-11-IDFI-023 (IIFR), ANR-18-IDEX-001 (UdP),  ANR-10-LABX-0083 (EFL).

%\vfill
%\pagebreak

\begin{small}
\bibliographystyle{IEEEbib}
\bibliography{main}
\end{small}
\end{document}